\documentclass{jors}
\usepackage{amsmath}
\usepackage{amssymb}
\usepackage{graphicx}

\definecolor{mygray}{gray}{0.6}

\begin{document}

\rule{\textwidth}{1pt}

\section*{(1) Overview}

\vspace{0.5cm}

\section*{Title}

%\textcolor{blue}{The title of the software paper should focus on the software, e.g. “Text mining software from the X project”. If the software is closely linked to a specific research paper, then “Software from Paper Title” is appropriate. The title should be factual, relating to the functionality of the software and the area it relates to rather than making claims about the software, e.g. “Easy-to-use”.}

An ObsPy library for event detection and seismic attribute calculation: preparing waveforms for automated analysis

\section*{Paper Authors}
%\textcolor{blue}{1. Last name, first name; (Lead/corresponding author first) \\
%2. Last name, first name; etc.}

1. Turner, Ross J;\\
2. Latto, Rebecca B;\\
3. Reading, Anya M\\

\section*{Paper Author Roles and Affiliations}
%\textcolor{blue}{1. First author role and affiliation \\
%2. Second author role and affiliation etc.}

1. School of Natural Sciences (Physics), University of Tasmania, Private Bag 37, Hobart, 7001, Australia; \\
2. School of Natural Sciences (Physics), University of Tasmania, Private Bag 37, Hobart, 7001, Australia; \\
3. School of Natural Sciences (Physics), University of Tasmania, Private Bag 37, Hobart, 7001, Australia; and Institute for Marine and Antarctic Studies, University of Tasmania, Private Bag 129, Hobart, TAS 7001, Australia. \\

\section*{Abstract}

%\textcolor{blue}{A short (ca. 100 word) summary of the software being described: what problem the software addresses, how it was implemented and architected, where it is stored, and its reuse potential.}

We have implemented an extension for the observational seismology \emph{obspy} software package to provide a streamlined tool tailored to the processing of seismic signals from non-earthquake sources, in particular those from deforming systems such as glaciers and landslides. This \emph{seismic attributes} library provides functionality to: (1) download and/or pre-process seismic waveform data; (2) detect and catalogue seismic events using multi-component signals from one or more seismometers; and (3) calculate characteristics (`attributes'/`features') of the identified events. The workflow is controlled by three main functions that have been tested for the breadth of data types expected from permanent and campaign-deployed seismic instrumentation. A selected STA/LTA-type (short-term average/long-term average), or other, event detection algorithm can be applied to the waveforms and user-defined functions implemented to calculate any required characteristics of the detected events. The code is written in Python 2/3 and is available on GitHub together with detailed documentation and worked examples.

\section*{Keywords}

%\textcolor{blue}{keyword 1; keyword 2; etc. \\
%Keywords should make it easy to identify who and what the software will be useful for.}

geophysics, seismology, data processing, reproducibility, Python

\section*{Introduction}

%\textcolor{blue}{An overview of the software, how it was produced, and the research for which it has been used, including references to relevant research articles. A short comparison with software which implements similar functionality should be included in this section. }

%- people normally visually classify events from active glaciers (e.g. whillians)\\
%- multi-sta/lta enables diverse events to be detected\\
%- pipeline in obspy not setup for easy reading/detection/extraction of events; assumes events pre-classified\\
%- attributes need to be determined to cluster events and separate signal types
%\\

Seismology provides an attractive tool to investigate physical processes in deforming systems. The seismic signals from active glaciers, for example, could enable monitoring of mechanisms including basal sliding \cite{Barcheck+2018, Lipovsky+2019, Winberry+2013}, fracturing \cite{Kavanaugh+2019}, melt water drainage \cite{Aso+2017}, and iceberg calving \cite{Nettles+2010, Olsen+2019}. The detection of seismic events from the recorded continuous seismic waveform data is a vital first step in any analysis. Event catalogues thus constructed are needed for local seismicity studies, comparisons between locations, or detection of change over extended time periods. However, the automated detection of seismic events is complicated in environmental and geotechnical seismology by the diverse populations of signal generation mechanisms. Those generated by active glaciers, for example, can be expected to span several orders of magnitude in duration and amplitude \cite{Podolskiy+2016}. Volcanoes \cite{Kohler+2010}, landslides \cite{Provost+2017} and mining activity \cite{Zhou+2018} similarly produce a broad range of seismic signals. As a result, the majority of cryoseismology studies to date use manual identification of events \cite{Barcheck+2018, Pomeroy+2013, Pratt+2014}. Manual techniques are not readily scalable nor exactly reproducible: in particular, they are not a first choice for monitoring applications nor for the data-driven detection of change.
\\

Established event detection algorithms have largely been developed for earthquake seismology \cite{Allen+1982, Allen+1978, Anstey+1966, Earle+1994}. These algorithms, including STA/LTA (short-term average/long-term average) \cite{Allen+1982} and template matching \cite{Anstey+1966}, are applied in real-time to seismic data to detect earthquakes and produce event catalogues \cite{Allen+1978, Earle+1994, Houliston+1984, Vaezi+2015}. The core classes in the \emph{obspy} software package are therefore designed assuming event metadata is available online alongside the waveform data \cite{obspycore+2020}; additional functions are included outside the core classes to select events directly from waveform data. These standard algorithms are generally only applicable to non-earthquake signals by employing an experimental approach to parameter selection \cite{Aster+2017, Podolskiy+2016}.
The \emph{seismic attributes} library provides software tools to download seismic waveform data from online repositories and detect environmental and geotechnical seismic events using a choice of algorithms. Algorithm options include the classic, recursive and delayed STA/LTA algorithms \cite{obspycore+2020}, and the newly developed multi-STA/LTA algorithm \cite{Latto+2020}. The multi-STA/LTA can simultaneously extract both short and long duration events of very different signal-to-noise levels and potentially enables real-time monitoring of cryoseismic events.
\\

Event catalogues constructed using the \emph{seismic attributes} library would be expected to comprise diverse signals generated by a range of mechanisms as noted above. The various signals typically need to be separated into related clusters based on the characteristics of their waveforms in order to study the events further \cite{Kohler+2010, Provost+2017}. For example, Provost et al. \cite{Provost+2017} consider 71 attributes based on waveform data in their study of landslide seismicity. These are broadly split into four categories: (1) waveform attributes (e.g. duration, energy, kurtosis); (2) spectral and spectrogram attributes (e.g. discrete Fourier transform); (3) network attributes (e.g. station with maximum amplitude); and (4) polarity attributes (e.g. azimuth, inclination).
The \emph{seismic attributes} library includes functions to calculate a number of standard signal properties: duration, ratio between ascending and descending time, energy in the autocorrelation function, energy in the frequency filtered spectrum, and the direction of wave propagation. These are provided in three bundles of attribute functions describing the waveform, spectrum and polarity of the signal; we do not include network attributes as these are unordered, discrete variables (and largely application dependent). User-defined functions can be added to derive customised characteristics within our software architecture. The correlation between attributes can be investigated using a plotting function to select the best subset for the subsequent application of (e.g.) clustering to a given set of events, informing the removal of redundant variables if appropriate. Data-driven techniques more generally, such as machine learning algorithms, may be readily applied to the calculated attributes to inform the current and future state of the glacier (i.e. identify signals in the lead-up to a large event \cite{Rouet-Leduc+2017}). 
\\
%some text moved to 'Reuse Potential'

\section*{Implementation and architecture}

%\textcolor{blue}{How the software was implemented, with details of the architecture where relevant. Use of relevant diagrams is appropriate. Please also describe any variants and associated implementation differences.}

The analysis of waveform data from environmental and geotechnical seismic deployments requires multiple distinct steps. Following the principle that one should write programs that do one thing well, and write programs that work together \cite{Salus+1994}, the \emph{seismic waveforms} library is correspondingly split into three primary functions to streamline the workflow. The name and purpose of these functions are as follows:
\begin{itemize}
\item \texttt{get\_waveforms()}; this function downloads waveform data from an online repository or alternatively reads data stored locally.
\item \texttt{get\_events()}; this function uses an event triggering algorithm to produce an event catalogue based on the seismic signals in one or more components of one or more seismic stations.
\item \texttt{get\_attributes()}; this function produces a \textit{pandas} \texttt{DataFrame} of the attributes for each event using functions included in the library, or user-defined attribute functions.
\end{itemize}
The library includes the numerous attribute functions that are called by \texttt{get\_attributes()}. Additional functions are also included to analyse the output of the primary functions (e.g. plotting), together with several private functions that support the primary workflow (documented in the code itself). The implementation of the three primary functions is outlined below.
\\

\textit{get\_waveforms}\\
Seismic datasets can build to a high data volume (several terrabytes) if high sampling rates are required over extended periods of time, or if using data from seismic arrays with multiple seismometers. This can result in memory problems when processing the waveforms and also disk space limitations. The first function in our workflow therefore acts as a gateway, to manage the volume of data downloaded from online repositories using \emph{obspy} \texttt{clients}. Waveform data from each seismometer is written into separate files comprising only a single day of data for a single component (e.g. vertical, north, east). The user can specify the write location of the files to split different time periods across several external drives. This same function checks if the requested data have previously been downloaded (by default) and can optionally not check for, or not download missing data (e.g. in the case of known data gaps). The waveform data are written as {.mseed} files using standard \emph{obspy} functions. 

The \texttt{get\_waveforms} function recombines downloaded or locally stored files into a single  \emph{obspy} \texttt{Trace} object for the requested time period, thus preparing the downloaded files for the next steps in the workflow. The user should examine small subsets of the time period of interest (e.g. a week at a time) based on factors including data volume and computer hardware performance. The seismic waveforms for the requested time period and components of a given seismometer are output as an \emph{obspy} \texttt{Stream} object. The streams from different seismometers are combined using the `+' operator into a single object.
\\

\textit{get\_events}\\
As noted in the introduction, several algorithms exist to trigger events from single component waveform data. The \texttt{get\_events} code first separates the signals from multiple seismometers and combines their (usually) three components into a single waveform. The user can optionally specify whether the component waveforms are combined as the sum-of-squares of their amplitude (i.e. Euclidean norm) to give the wave energy, or the absolute value of the wave amplitude. Further, to ensure that taking the absolute value of the amplitude does not affect the results (e.g. doubling frequency) we tested a computationally intensive option to fit the (time-varying) principle component of the direction of wave oscillation to obtain a signed amplitude; this gave identical event detections to the absolute amplitude and so is not included in our published version of the code due to the significantly longer computation time.

\begin{figure}
\includegraphics[width=\textwidth]{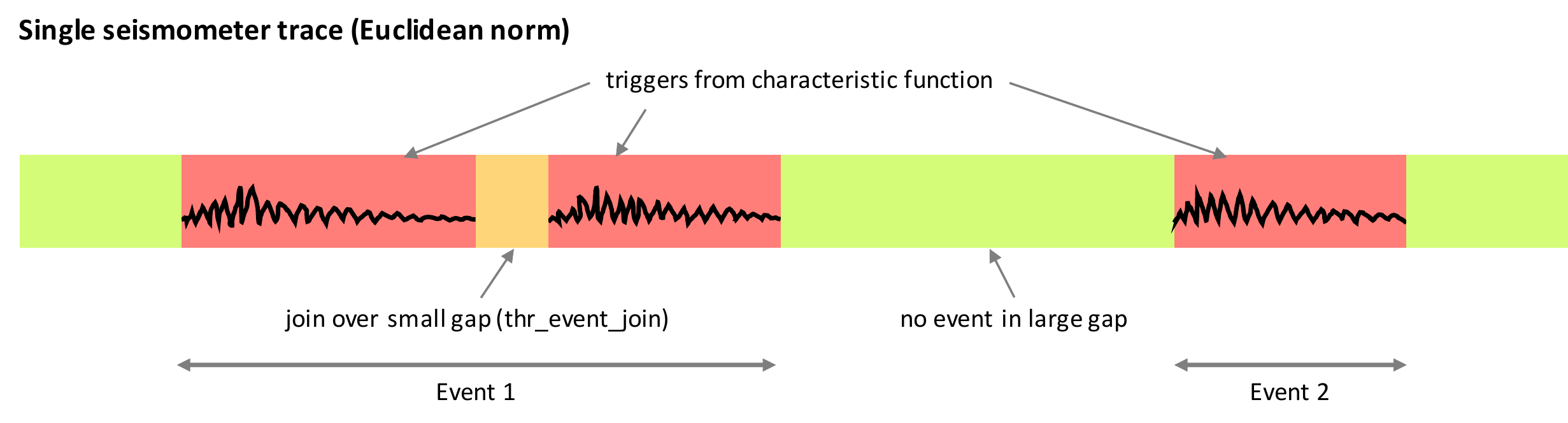}\vspace{8pt}
\includegraphics[width=\textwidth]{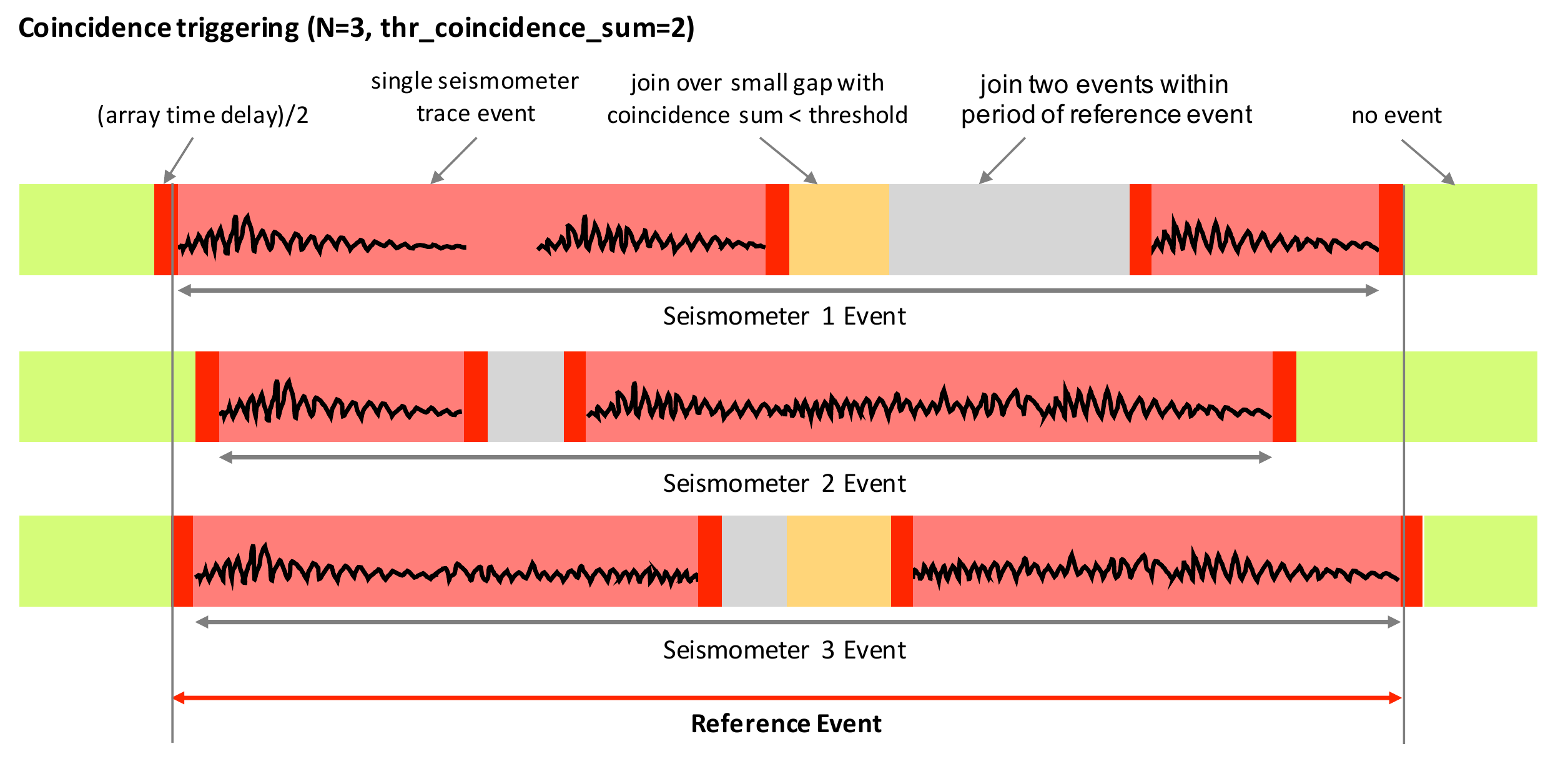}
\centering
\caption{\textit{Top:} Diagrammatic representation of algorithm used to identify events from a single seismometer comprising one or more channels (traces show their Euclidean norm). The characteristic function for an STA/LTA algorithm is used to `trigger' events (shown in red) and small gaps between these events (shorter than \texttt{thr\_event\_join}) are ignored (orange). No events are present at other times (shown in green). \textit{Bottom:} Representation of algorithm used to identify events in the `reference event' and `trace' catalogues for seismic arrays (with multiple seismometers). The example shows an indicative array with $N=3$ seismometers and \texttt{thr\_coincidence\_sum} $\geqslant 2$ simultaneous detections. Events identified for the single seismometers traces are shown in red (as given in the top panel). The duration of these events is extended at each end by half the delay in arrival time of a wave between the most distant seismometers in the array (shown in deep red). The reference event is identified as the times when \texttt{thr\_coincidence\_sum} $\geqslant 2$ seismometers have a detection (shown in red or deep red); small gaps with fewer seismometers are joined over (shown in orange). The events at each seismometer are simply the events identified from the single seismometer traces (red but not deep red), but with any times between events that occur during the reference event also included (shown in grey).}
\label{fig:trigger}
\end{figure}

We use an adapted version of the \emph{obspy} \texttt{coincidence\_trigger} function to create a reference catalogue of events based on the STA/LTA characteristic function at each seismometer for their combined component waveforms. Our version of the \texttt{coincidence\_trigger} function includes adjustments in the algorithm to better align with the outputs needed in our workflow and improvements to computational performance that are possible for our narrower use of this function (see Figure \ref{fig:trigger} for schematic of algorithm). This function can use the standard STA/LTA algorithms available within \textit{obspy} \cite{obspycore+2020}, and optionally the recently developed multi-STA/LTA algorithm \cite{Latto+2020}. In all cases, location-specific algorithm parameters are likely to be more successful for environmental seismology applications. The start and stop times of the detected event from a single seismometer can be extended to consider small gaps in between triggers (up to length \texttt{thr\_event\_join}), creating a single, longer-duration record. The reference event catalogue is created by finding times when $n$ (i.e. \texttt{thr\_coincidence\_sum}) of the $N$ seismometers in the array have temporally coincident records. The physical dimensions of the seismic array are calculated at this step to estimate an upper bound on the delay in arrival time of the signal at different seismometers. This delay is added to the duration of the single seismometer records to ensure the reference events are not artificially shortened. The reference event catalogue, which includes a reference start time and duration for each event, is output as a \emph{pandas} \texttt{DataFrame} to provide a convenient format to write to file and interrogate.

\begin{figure}
\includegraphics[width=0.95\textwidth]{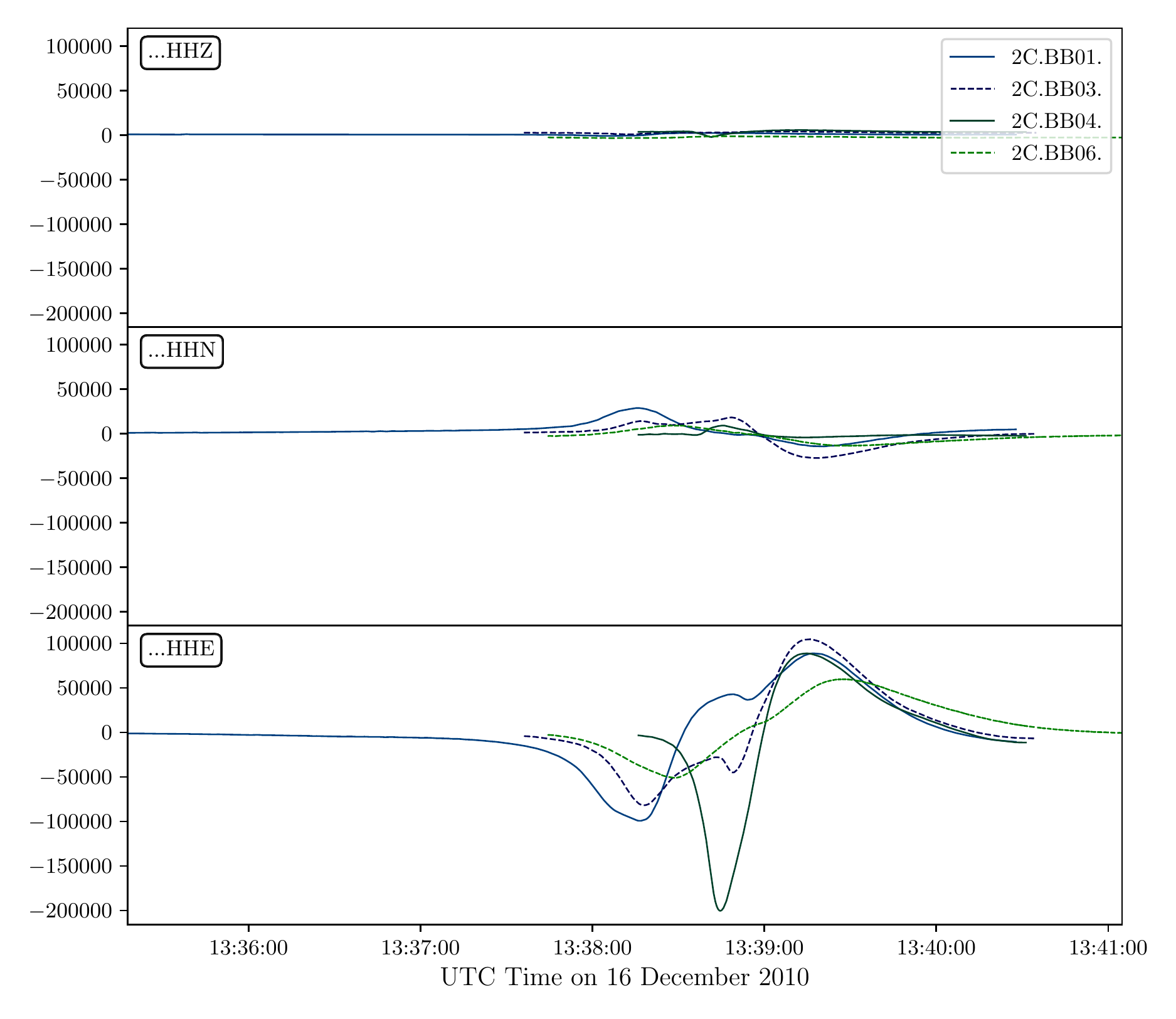}
\centering
\caption{Waveforms of an event detected using four seismometers (BB01, BB03, BB04 and BB06) comprising part of a seismic array on the Whillans Ice Stream in Antarctica from 13:35:37 on 16 December 2010. The top, middle and bottom panels show the vertical, north and east component of the signal respectively. The waveforms for each seismometer start 30 seconds prior to the start time in the trace catalogue and terminate 60 seconds after the stop time. The code to reproduce this plot is provided as a worked example.}
\label{fig:waveforms}
\end{figure}

Our adapted version of the \texttt{coincidence\_trigger} function provides additional utility for other researchers by outputting a secondary catalogue of trace metadata, for each seismometer, for all identified reference events. The triggered records at each seismometer that occur during (at least) part of the time period covered by a given reference event are joined together (if not already a single record). This record may extend beyond the time period of the reference event, be contained entirely within it, or have no detection at all in the case of weaker events. This trace (metadata) catalogue, which includes the trigger start time and duration for all seismometers, is similarly output as a \emph{pandas} \texttt{DataFrame}. The trace metadata can be used to extract waveform data for identified events in the catalogue in a format that will be familiar to seismologists, for example, as shown in Figure \ref{fig:waveforms} for a low-frequency event detected using four seismometers on the Whillans Ice Stream in Antarctica from 13:35:37 on 16 December 2010.
\\

\begin{figure}
\includegraphics[width=0.95\textwidth]{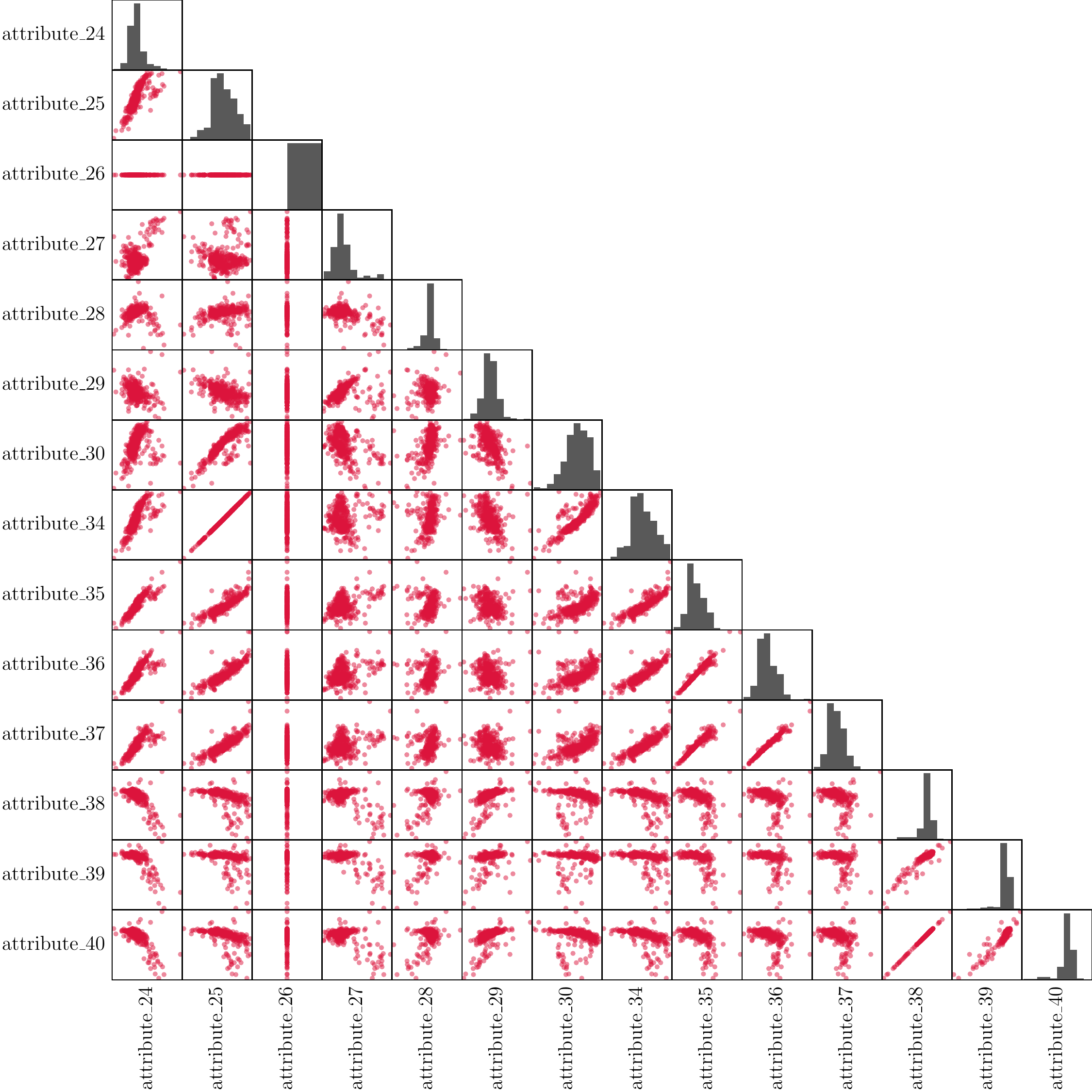}
\centering
\caption{Corner (or pair) plot illustrating the correlation between spectral attributes for the events detected at Ilulissat, Greenland on 1 January 2018 using the recursive STA/LTA algorithm. The attribute names are defined to match those used by Provost et al. \cite{Provost+2017}. In this example, attribute 26 considers a frequency range close to the sampling rate of the recorded signal, therefore, it has null value for all events. The code to reproduce this plot is provided as a worked example in the detailed documentation provided with the software.}
\label{fig:correlation}
\end{figure}

\newpage
\textit{get\_attributes}\\
The \texttt{get\_attributes} function is written following further principles \cite{Salus+1994}, that one should write flexible and open programs. This function extracts the waveform data for a given seismometer for the duration of a given event listed in the reference event, or trace (metadata) catalogue. The waveforms for each component are stored as separate \texttt{Trace} objects in a single stream. This \texttt{Stream} object is passed to one or multiple attribute functions to derive the characteristics of the given event as measured by a given seismometer. The values of the requested attributes for the array of seismometers are output as a \emph{pandas} \texttt{DataFrame}. The correlation between the spectral attributes included in the library are shown in Figure \ref{fig:correlation} for events detected at Ilulissat, Greenland (DK.ILULI) on 1 January 2018 using the recursive STA/LTA algorithm.

Optionally, custom attribute functions may be added to calculate any chosen characteristic of a waveform. Custom functions must take a \texttt{Stream} object containing one or more components as an input, and output the attribute name and value. Public functions to combine the component waveforms into the wave energy or absolute value of the wave amplitude are included in the \emph{seismic attributes} library to aid the user. The attribute functions (as many as are required) are passed to the \texttt{get\_attributes} function as optional parameters, with return values stored as above. 

\section*{Quality control}

%\textcolor{blue}{Detail the level of testing that has been carried out on the code (e.g. unit, functional, load etc.), and in which environments. If not already included in the software documentation, provide details of how a user could quickly understand if the software is working (e.g. providing examples of running the software with sample input and output data). }

We have tested the \textit{seismic attributes} library to find any software bugs and to ensure outputs are reproducible by other researchers running the code or using different platforms \cite{LeVeque+2012}. We first created test cases based on glacier seismic signals from a field campaign seismic array on the Whillans Ice Stream in West Antarctica \cite{Winberry+2013}. Plausible uses of the code were tested by selecting some or all of the seismometers in the array, some or all of the three components of the signal recorded at each seismometer (i.e. vertical, north and east), and considering different durations of requested time (with and without small data gaps). The optional function inputs were also tested in this manner. In application, e.g. \cite{Latto+2020}, the event catalogue produced using the \emph{seismic attributes} workflow compared favourably to events that were visually classified using the same set of seismometers \cite{Pratt+2014}. Finally, the attribute functions included in the library were tested on mock waveform data to ensure they produce the theoretically expected results. 

Testing was primarily carried out using a Python 3.8 installation on a macOS 10/11 system (Unix). The key features of the \emph{seismic attributes} library were tested on other platforms for compatability, including both Python 2.7 and 3.7/8 for the most up to date version of \emph{obspy}, \emph{numpy}, \emph{pandas} and \emph{seaborn} available for that installation. The code was further tested on both macOS 10/11 and Windows operating systems, especially to verify functionality of the different file systems. The reference event and trace (metadata) catalogues were compared between these platforms to assess the consistency of code functionality.

\section*{(2) Availability}
\vspace{0.5cm}
\section*{Operating system}

macOS 10/11 (including Unix and Linux) and Windows.

\section*{Programming language}

Python 2 or 3.

\section*{Additional system requirements}

Memory and disk space will limit the volume of data that can be processed in a contiguous chunk; the code will work on any system that can support \emph{obspy}. Internet access is required to download new waveform data from online repositories.

\section*{Dependencies}

The minimal dependency for use is Python 2/3 with \emph{obspy}, \emph{numpy}, \emph{pandas} and \emph{seaborn} packages installed.

%\section*{List of contributors}

%\textcolor{blue}{Please list anyone who helped to create the software (who may also not be an author of this paper), including their roles and affiliations.}

\section*{Software location:}

{\bf Archive} %\textcolor{blue}{(e.g. institutional repository, general repository) (required – please see instructions on journal website for depositing archive copy of software in a suitable repository)} 

\begin{description}[noitemsep,topsep=0pt]
	\item[Name:] GitHub
	\item[Persistent identifier:] \url{https://github.com/rossjturner/seismic\_attributes}
	\item[Licence:] GPL version 3
	\item[Publisher:] Ross James Turner
	\item[Date published:] 14/10/2020
\end{description}

{\bf Code repository} %\textcolor{blue}{(e.g. SourceForge, GitHub etc.) (required)}

\begin{description}[noitemsep,topsep=0pt]
	\item[Name:] GitHub
	\item[Persistent identifier:] \url{https://github.com/rossjturner/seismic\_attributes}
	\item[Licence:] GPL version 3
	\item[Date published:] 14/10/2020
\end{description}

\section*{Language}

Python 2 or 3

\section*{(3) Reuse potential}

%\textcolor{blue}{Please describe in as much detail as possible the ways in which the software could be reused by other researchers both within and outside of your field. This should include the use cases for the software, and also details of how the software might be modified or extended (including how contributors should contact you) if appropriate. Also you must include details of what support mechanisms are in place for this software (even if there is no support).}

The main purpose behind the \emph{seismic attributes} library is to provide a streamlined workflow for researchers to detect and categorise seismic events from environmental and geotechnical sources in a rigorous and reproducible manner. The code will find extensive use in areas of seismology where a diverse population of seismic signals are present, including glaciers, volcanoes, landslides and mine sites. The \texttt{get\_events} function is expected to be especially useful, in particular in glacier seismology, to provide a consistent method for creating event catalogues for ongoing machine learning and conventional seismological analysis.  We anticipate that our \texttt{get\_attributes} function may also be useful in applications outside seismology to aid in the construction of a catalogue of waveform attributes for use in machine learning.  The code would need only minor modifications to handle time series data, outside of seismology, stored in different formats. It could conceivably find applications other areas of science (e.g. astrophysics), economics and many other potential applications. Limited support may be provided by contacting the corresponding author.

%\section*{Acknowledgements}

\section*{Funding statement}

%\textcolor{blue}{If the software resulted from funded research please give the funder and grant number.}

This research was supported under Australian Research Council’s (ARC) Special Research Initiative for Antarctic Gateway Partnership (project ID SR140300001), and ARC Discovery Projects DP190100418 and DP2101000834. 

\section*{Competing interests}

The authors declare that they have no competing interests.

%\section*{References}

%\textcolor{blue}{Please enter references in the Harvard style and include a DOI where available, citing them in the text with a number in square brackets, e.g. \\ }

%\begin{thebibliography}{9}
\bibliographystyle{abbrv}
\bibliography{seismic_attributes.bib}
%\bibitem{Latto+2020}
%Latto, R B, Turner, R J and Reading, A M (2020). Event detection for cryoseismology. Cryosphere. in press

%\bibitem{Latto+2021}
%Latto, R B, Turner, R J and Reading, A M (2021). Machine learning paper
%\end{thebibliography}

\vspace{2cm}

\rule{\textwidth}{1pt}

{ \bf Copyright Notice} \\
Authors who publish with this journal agree to the following terms: \\

Authors retain copyright and grant the journal right of first publication with the work simultaneously licensed under a  \href{http://creativecommons.org/licenses/by/3.0/}{Creative Commons Attribution License} that allows others to share the work with an acknowledgement of the work's authorship and initial publication in this journal. \\

Authors are able to enter into separate, additional contractual arrangements for the non-exclusive distribution of the journal's published version of the work (e.g., post it to an institutional repository or publish it in a book), with an acknowledgement of its initial publication in this journal. \\

By submitting this paper you agree to the terms of this Copyright Notice, which will apply to this submission if and when it is published by this journal.

\end{document}